\documentclass{article}
\usepackage{epsfig}
\usepackage{amsmath}

\input{tcilatex}

\begin{document}

\title{Entanglement-Assisted Classical Capacity of Quantum Channels with
Correlated Noise}
\author{Nigum Arshed\thanks{%
nigum@phys.qau.edu.pk} and A. H. Toor\thanks{%
ahtoor@qau.edu.pk} \and Department of Physics, Quaid-i-Azam University \and %
Islamabad 45320, Pakistan. }
\maketitle

\begin{abstract}
We calculate the entanglement-assisted classical capacity of symmetric and
asymmetric Pauli channels where two consecutive uses of the channels are
correlated. It is evident from our study that in the presence of memory, a
higher amount of classical information is transmitted over quantum channels
if there exists prior entanglement as compared to product and entangled
state coding.
\end{abstract}

Unlike classical channels, more than one distinct capacities are associated
with quantum channels \cite{BSST}, depending on the type of information
(classical or quantum) transmitted and the additional resources brought into
play. Calculating the capacities of quantum channels is an important task of
quantum information theory. Most of the work, so far, has focused on
memoryless quantum channels \cite{BSST}, \cite{HSW}. A channel is memoryless
if noise acts independently over each use of the channel. In practice, the
noise in consecutive uses of the channel is not independent and exhibits
some correlation. The correlation strength is determined by the degree of
memory of the channel.

Quantum channels with memory were considered recently by Macchiavello and
Palma \cite{Chiara-1}. They studied the depolarizing channel with Markov
correlated noise and showed that beyond a certain threshold in the degree of
memory of the channel, coding with maximally entangled states has edge over
product states. Later this work was extended to the case of a non-Pauli
channel and similar behavior was reported \cite{Yeo}. An upper bound for the
maximum mutual information of quantum channels with partial memory was given
by Macchiavello \textit{et al.\ }\cite{Chiara-2}. This bound is achieved for
minimum entropy states which turned out to be entangled above the memory
threshold and proved that entangled states are optimal for the transmission
of classical information over quantum channels. The upper bounds for the
classical information capacity of indecomposable quantum memory channels %
\cite{Bowen-2}, and the asymptotic classical and quantum capacities of
finite memory channels \cite{Bowen-1}, were also derived recently.

Quantum memory channel can be modeled as a unitary interaction between the
states transmitted through the channel, independent environment and the
channel memory state that remains unchanged during the interaction \cite%
{Bowen-1}. An experimental model for quantum channels with memory motivated\
by random birefringence fluctuations in a fibre optic link was recently
proposed \cite{Ball}, and demonstrated experimentally \cite{Konrad}. It was
inferred in both the studies that entanglement is a useful resource to
enhance the classical information capacity of quantum channels. A general
model for quantum channels with memory was presented in Ref. \cite{Werner}.
It was shown that under mild causality constraints every quantum process can
be modeled as a concatenated memory channel with some memory initializer.

In this paper, we calculate the entanglement-assisted classical capacity of
Pauli channels with correlated noise. Our results show that provided the
sender and receiver share prior entanglement, a higher amount of classical
information is transmitted over Pauli channels (in the presence of memory)
as compared to product and entangled state coding.

We begin with a brief description of quantum memory channels. Quantum
channels model the noise that occur in an open quantum system due to
interaction with the environment. Mathematically, a quantum channel $%
\mathcal{N}$ is defined as a completely positive, trace preserving map from
input state density matrices to output state density matrices. If the state
input to the channel is $\rho $\ then in Kraus representation \cite{Preskill}%
, action of the channel is described as 
\begin{equation}
\mathcal{N}\left( \rho \right) =\sum_{k}E_{k}\rho E_{k}^{\dagger },
\label{Kraus-representation}
\end{equation}%
where $E_{k}$ are the Kraus operators of the channel which satisfy the
completeness relationship, i.e., $\sum_{k}E_{k}^{\dagger }E_{k}=I$. Here we
restrict ourselves to Pauli channels that map identity to itself, that is, $%
\mathcal{N}\left( I\right) =I$.

The action of a quantum channel $\mathcal{N}$ on the input state density
matrix $\rho _{n}$, consisting of $n$ qubits (including entangled ones) is
given by

\begin{equation}
\mathcal{N}\left( \rho _{n}\right) =\sum_{k_{1}\cdots k_{n}}p_{k_{1}\cdots
k_{n}}\left( E_{k_{n}}\otimes \cdots \otimes E_{k_{1}}\right) \rho
_{n}\left( E_{k_{n}}^{\dagger }\otimes \cdots \otimes E_{k_{1}}^{\dagger
}\right) ,  \label{memoryless}
\end{equation}%
where the Kraus operators $E_{k_{n}}\otimes \cdots \otimes E_{k_{1}}$ are
applied with probability $p_{k_{1}\cdots k_{n}}$ which satisfies $%
\sum_{k_{1}\cdots k_{n}}$ $p_{k_{1}\cdots k_{n}}$ $=1$. The quantity $%
p_{k_{1}\cdots k_{n}}$ can be interpreted as the probability that a random
sequence of operations is applied to the sequence of $n$ qubits transmitted
through the channel. For a memoryless channel, these operations are
independent therefore, $p_{k_{1}\cdots k_{n}}=p_{k_{1}}p_{k_{2}}\cdots
p_{k_{n}}$. In the presence of memory they exhibit some correlation. A
simple example is given by the Markov chain, i.e.,%
\begin{equation}
p_{k_{1}\cdots k_{n}}=p_{k_{1}}p_{k_{2}\mid k_{1}}\cdots p_{k_{n}\mid
k_{n-1}}.  \label{Markov}
\end{equation}%
In the above expression, $p_{k_{n}\mid k_{n-1}}$ is the conditional
probability that an operation, say $E_{k_{n}}$, is applied to the $n$th
qubit provided that it was applied on the $\left( n-1\right) $th qubit. The
Kraus operators for two consecutive uses of a Pauli channel with partial
memory are \cite{Chiara-1} 
\begin{equation}
E_{i,j}=\sqrt{p_{i}\left[ \left( 1-\mu \right) p_{j}+\mu \delta _{i,j}\right]
}\sigma _{i}\otimes \sigma _{j},\text{ \ \ }0\leq \mu \leq 1
\label{partial-memory}
\end{equation}%
where $\mu $ is the memory coefficient of the channel and $\sigma _{i,j}$,
where $i,j=0,x,y,z$ are the Pauli operators with $\sigma _{0}=I$. It is
evident from the above expression that the same operation is applied to both
qubits with probability $\mu $ while with probability $1-\mu $ both
operations are uncorrelated.

Entanglement, a fundamental resource of quantum information theory, can be
used to enhance the classical capacity of quantum channels in two different
ways. One, by encoding classical information on entangled states and two, by
sharing prior entanglement between the sender and receiver. For noiseless
quantum channels, the classical capacity is doubled if there exists prior
entanglement, i.e., $C_{E}=2C$ \cite{densecoding}. The amount of classical
information transmitted reliably through a noisy quantum channel $\mathcal{N}
$ \ with the help of unlimited prior entanglement is given by its
entanglement-assisted classical capacity \cite{BSST}

\begin{equation}
C_{E}(\mathcal{N})=\max_{\rho \in \mathcal{H}_{in}}S(\rho )+S(\mathcal{N}%
(\rho ))-S((\mathcal{N}\otimes \mathcal{I})\Phi _{\rho }),  \label{CE}
\end{equation}%
which is, the maximum over the input distribution of the input-output
quantum mutual information. In the above expression%
\begin{equation}
S(\rho )=-\text{Tr}\left( \rho \log _{2}\rho \right) ,  \label{von-Neumann}
\end{equation}%
is the von Neumann entropy of the input state density matrix $\rho $, $%
S\left( \mathcal{N}\left( \rho \right) \right) $ is the von Neumann entropy
of the output state density matrix and $S(\left( \mathcal{N}\otimes \mathcal{%
I}\right) \Phi _{\rho })$ is the von Neumann entropy of the purification $%
\Phi _{\rho }\in \mathcal{H}_{in}\otimes \mathcal{H}_{ref}$ of $\rho $ over
a reference system $\mathcal{H}_{ref}$. The maximally entangled state $\Phi
_{\rho }$ shared by the sender and receiver provides a purification of the
input state $\rho $. Half of the purification Tr$_{\text{ref}}\Phi _{\rho
}=\rho $ is transmitted through the channel $\mathcal{N}$ while the other
half $\mathcal{H}_{ref}$ is sent through the identity channel $\mathcal{I}$
(this corresponds to the portion of the entangled state that the receiver
holds at the start of the protocol. See Fig. 1 in Ref. \cite{BSST}).

In the following we calculate the entanglement-assisted classical capacity
of some well known Pauli channels for two consecutive uses of the channels.
The channels considered are assumed to have partial memory. Suppose that the
sender $A$ and receiver $B$ share two (same or different) maximally
entangled Bell states\footnote{%
Entanglement is an interconvertable resource. It can be transformed
(concentrated or diluted) reversibly, with arbitrarily high fidelity\ and
asymptotically negligible amount of classical communication \cite{Bennett}, %
\cite{Lo}. Therefore, it is sufficient to use Bell states as entanglement
resource in calculating $C_{E}$.}, i.e., 
\begin{subequations}
\begin{eqnarray}
\left| \psi ^{\pm }\right\rangle &=&\frac{1}{\sqrt{2}}\left( \left|
00\right\rangle \pm \left| 11\right\rangle \right) ,  \label{bell-1} \\
\left| \phi ^{\pm }\right\rangle &=&\frac{1}{\sqrt{2}}\left( \left|
01\right\rangle \pm \left| 10\right\rangle \right) .  \label{bell-2}
\end{eqnarray}%
The first qubit of the Bell states belongs to the sender while the second
qubit belongs to the receiver. As only sender's qubits pass through the
channel, therefore, the input state density matrix $\rho $ is obtained by
performing trace over the receiver's system. 
\end{subequations}
\begin{equation}
\rho =\text{Tr}_{B}\left( \left| \psi ^{+}\right\rangle \left\langle \psi
^{+}\right| \right) \otimes \text{Tr}_{B}\left( \left| \psi
^{-}\right\rangle \left\langle \psi ^{-}\right| \right) =\frac{I}{4},
\label{input}
\end{equation}%
where $I$ is the $4\times 4$ identity matrix. The purification $\Phi _{\rho
} $ of the input state $\rho $ is 
\begin{eqnarray}
\Phi _{\rho } &=&\left| \psi ^{+}\right\rangle \left\langle \psi ^{+}\right|
\otimes \left| \psi ^{-}\right\rangle \left\langle \psi ^{-}\right|  \notag
\\
&=&\frac{1}{16}[\left( \sigma _{00}+\sigma _{zz}\right) \otimes \left\{
\left( \sigma _{00}+\sigma _{zz}\right) -\left( \sigma _{xx}-\sigma
_{yy}\right) \right\}  \notag \\
&&+\left( \sigma _{xx}-\sigma _{yy}\right) \otimes \left\{ \left( \sigma
_{00}+\sigma _{zz}\right) -\left( \sigma _{xx}-\sigma _{yy}\right) \right\}
],  \label{purification}
\end{eqnarray}%
with $\sigma _{ii}=\sigma _{i}\otimes \sigma _{i}$. The purification $\Phi
_{\rho }$ is the joint state of the sender and receiver as explained
previously. Using the definition of Pauli channels, it is straight forward
to write the output state density matrix $\mathcal{N}\left( \rho \right) $
as 
\begin{equation}
\mathcal{N}\left( \rho \right) =\sum_{i,j}E_{i,j}\rho E_{i,j}^{\dagger }=%
\frac{I}{4},  \label{output}
\end{equation}%
with $E_{i,j}$ given by the Eq. (\ref{partial-memory}). Therefore, for
symmetric and asymmetric Pauli channels%
\begin{equation}
S\left( \rho \right) +S\left( \mathcal{N}(\rho )\right) =4.
\label{first-terms}
\end{equation}%
However, the transformation of the purification $\Phi _{\rho }$ under the
action of Pauli channels is different for all channels. In the presence of
partial memory, the action of Pauli channels on the purification $\Phi
_{\rho }$ is described by the Kraus operators

\begin{equation}
\widetilde{E}_{i,j}=\sqrt{p_{i}\left[ \left( 1-\mu \right) p_{j}+\mu \delta
_{i,j}\right] }(\sigma _{i}\otimes I)\otimes (\sigma _{j}\otimes I).
\label{Kraus-Purification}
\end{equation}%
Von Neumann entropy of the purification state transformed under the action
of Pauli channels is given by%
\begin{equation}
S((\mathcal{N}\otimes I)\Phi _{\rho })=-\sum_{i}\lambda _{i}\log _{2}\lambda
_{i}.  \label{Third Term}
\end{equation}%
where $\lambda _{i}$ are the eigenvalues of the transformed purification
state. Now we consider some examples of Pauli channels, both symmetric and
asymmetric, and work out their entanglement-assisted classical capacity.

The depolarizing channel is a Pauli channel with particularly nice symmetry
properties \cite{qcqi}. In the presence of partial memory, the action of
depolarizing channel on the purification $\Phi _{\rho }$ is described by the
Kraus operators $\widetilde{E}_{i,j}$ with $i,j=0,x,y,z$. Pauli operators $%
\sigma _{i,j}$, given in the Eq. (\ref{Kraus-Purification}) are applied with
probabilities $p_{0}=\left( 1-p\right) ,$ $p_{x}=$ $p_{y}=$ $p_{z}=\frac{p}{3%
}$. Eigenvalues of the purification $\Phi _{\rho }$ transformed under the
action of the depolarizing channel are 
\begin{subequations}
\begin{eqnarray}
\lambda _{1}^{D} &=&\frac{1}{16}\left( 1+3\eta \right) \left\{ \left(
1+3\eta \right) \left( 1-\mu \right) +4\mu \right\} ,  \label{dp-eigen1} \\
\lambda _{2,\cdots ,7}^{D} &=&\frac{1}{16}\left( 1-\eta \right) \left(
1+3\eta \right) \left( 1-\mu \right) ,  \label{dp-eigen2} \\
\lambda _{8,9,10}^{D} &=&\frac{1}{16}\left( 1-\eta \right) \left\{ \left(
1-\eta \right) \left( 1-\mu \right) +4\mu \right\} ,  \label{dp-eigen3} \\
\lambda _{11,\cdots ,16}^{D} &=&\frac{1}{16}\left( 1-\eta \right) ^{2}\left(
1-\mu \right) ,  \label{dp-eigen4}
\end{eqnarray}%
where $\eta =1-\frac{4}{3}p$, is the shrinking factor for single use of
depolarizing channel. The entanglement-assisted classical capacity of the
depolarizing channel in the presence of partial memory is the sum of Eqs. (%
\ref{first-terms}) and (\ref{Third Term}), with $\lambda _{i}$ given by Eqs.
(\ref{dp-eigen1})-(\ref{dp-eigen4}). For $0\leq \eta \leq 1$, $\log _{2}\eta
<0$, which makes the term given by Eq. (\ref{Third Term}) negative and
reduces the capacity of memoryless depolarizing channel below the factor of $%
4$ (i.e., $C_{E}$\ for two uses of the noiseless depolarizing channel). As
the degree of memory $\mu $ of the channel increases, the factor $\left(
1-\mu \right) \rightarrow 0$ which makes the contribution of error term
i.e., Eq. (\ref{Third Term}), small. Therefore, we conclude that memory of
the channel increases the entanglement-assisted classical capacity of
depolarizing channel.

\begin{figure}[h]
\centerline{
\epsfig{file=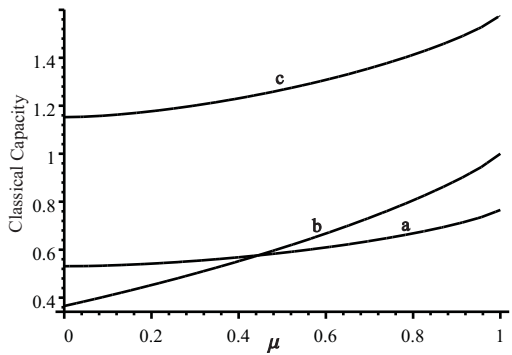}
}
\caption{Plot of classical capacity $C$ (for product (a) and entangled (b)
state coding) and entanglement-assisted classical capacity $C_E$ versus the
memory coefficient $\protect\mu$ for depolarizing channel, with $\protect\eta%
=0.8$. The capacities are normalized with respect to the number of channel
uses.}
\end{figure}

Figure 1 gives the plot of the classical capacity $C$ and the
entanglement-assisted classical capacity $C_{E}$ of depolarizing channel
versus its memory coefficient $\mu $. As reported in Ref. \cite{Chiara-1}
beyond a certain memory threshold entangled states enhance the classical
information capacity of the channel. It is evident from Fig.1 that a higher
amount of classical information is transmitted if prior entanglement is
shared by the sender and receiver. We infer that prior entanglement has
clear edge over both product and entangled state coding for all values of $%
\mu $.

Next we consider some examples of asymmetric Pauli channels. The simplest
example is given by the Flip channels \cite{qcqi}. The noise introduced by
them is of three types namely, bit flip, phase flip and bit phase flip.
Kraus operators of\textit{\ }flip channels with partial memory acting on the
purification $\Phi _{\rho }$ are given by the Eq. (\ref{Kraus-Purification}%
), with $i,j=0,f$,\emph{\ }applied with probabilities $p_{0}=\left(
1-p\right) ,$ $p_{f}=p$. Here $f=x,z$ and $y$, for bit flip, phase flip and
bit-phase flip channels, respectively. The purification $\Phi _{\rho }$ is
mapped by the flip channels to an output state purification having
eigenvalues 
\end{subequations}
\begin{subequations}
\begin{eqnarray}
\lambda _{1}^{F} &=&\frac{1}{4}\left( 1+\chi \right) \left\{ 1+\mu +\chi
\left( 1-\mu \right) \right\} ,  \label{f-eigen1} \\
\lambda _{2,3}^{F} &=&\frac{1}{4}\left( 1-\chi ^{2}\right) \left( 1-\mu
\right) ,  \label{f-eigen2} \\
\lambda _{4}^{F} &=&\frac{1}{4}\left( 1-\chi \right) \left\{ 1+\mu -\chi
\left( 1-\mu \right) \right\} ,  \label{f-eigen3} \\
\lambda _{5,\cdots ,16}^{F} &=&0,  \label{f-eigen4}
\end{eqnarray}%
where $\chi =1-2p$ is the shrinking factor of flip channels, for single use
of the channels. The entanglement-assisted classical capacity for flip
channels with partial memory over two consecutive uses of the channels is
given by the sum of Eqs. (\ref{first-terms}) and (\ref{Third Term}). The
Eqs. (\ref{f-eigen1})-(\ref{f-eigen4}) give $\lambda _{i}$ for flip channels.

Secondly, consider the two-Pauli channel. The Kraus operators $\widetilde{E}%
_{i,j}$ for two-Pauli channel with partial memory acting on the purification
are given by the Eq. (\ref{Kraus-Purification}). For two-Pauli channel, $%
i,j=0,x,y$, and the Pauli operators $\sigma _{i,j}$ are applied with
probabilities $p_{0}=\left( 1-p\right) ,$ $p_{x}=p_{y}=\frac{p}{2}$. The
purification $\Phi _{\rho }$ transformed under the action of two-Pauli
channel has eigenvalues 
\end{subequations}
\begin{subequations}
\begin{eqnarray}
\lambda _{1}^{TP} &=&\zeta _{1}\left\{ \zeta _{1}\left( 1-\mu \right) +\mu
\right\} ,  \label{tp-eigen1} \\
\lambda _{2,\cdots ,5}^{TP} &=&\frac{1}{2}\zeta _{1}\left( 1-\zeta
_{1}\right) \left( 1-\mu \right) ,  \label{tp-eigen2} \\
\lambda _{6,7}^{TP} &=&\frac{1}{4}\left( 1-\zeta _{1}\right) \left\{ 1+\mu
-\zeta _{1}\left( 1-\mu \right) \right\} ,  \label{tp-eigen3} \\
\lambda _{8,9}^{TP} &=&\frac{1}{4}\left( 1-\zeta _{1}\right) ^{2}\left(
1-\mu \right) ,  \label{tp-eigen4} \\
\lambda _{10,\cdots ,16}^{TP} &=&0.  \label{tp-eigen5}
\end{eqnarray}%
For single use of the two-Pauli channel, $\zeta _{1}=1-p$ is the shrinking
factor for the states $\sigma _{x}$ and $\sigma _{y}$\ while the shrinking
factor for $\sigma _{z}$ is $\zeta _{2}=1-2p$. In the presence of partial
memory, the entanglement-assisted classical capacity of two-Pauli channel is
the sum of Eqs. (\ref{first-terms}) and (\ref{Third Term}), where $\lambda
_{i}$ are given by Eqs. (\ref{tp-eigen1})-(\ref{tp-eigen5}).

Finally, we consider the phase damping channel \cite{qcqi}. In the presence
of partial memory, Kraus operators of phase damping channel acting on the
purification are given by the Eq. (\ref{Kraus-Purification}), with $i,j=0,z$%
, applied with probabilities $p_{0}=\left( 1-\frac{p}{2}\right) ,$ $p_{z}=%
\frac{p}{2}$. The expression of entanglement-assisted classical capacity of
phase damping channel is identical to that for flip channels, with $\chi $
replaced by $\gamma =1-p$ which is the shrinking factor for single use of
the phase damping channel.

\begin{figure}[h]
\centerline{
\epsfig{file=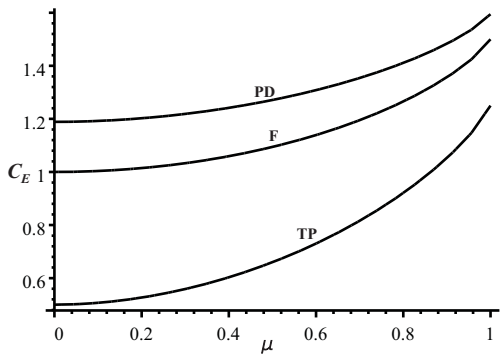}
}
\caption{Plot of entanglement-assisted classical capacity $C_E$ and memory
coefficient $\protect\mu$ for flip channels (f), phase damping channel (pd)
and two-Pauli channel (tp), for $p=0.5$. The capacities are normalized with
respect to the number of channel uses.}
\end{figure}

Figure 2 gives the plot of the entanglement-assisted classical capacity $%
C_{E}$ versus the\ memory coefficient $\mu $ for flip channels, two-Pauli
channel and phase damping channel. It is evident from the plot that the
capacity increases continuously with the degree of memory of the channels
and for a given error probability $p$ acquires its maximum value for $\mu =1$%
, i.e., perfect memory.

In conclusion, in this paper we have calculated the entanglement-assisted
classical capacity of symmetric and asymmetric Pauli channels in the
presence of memory. The noise in two consecutive uses of the channels is
assumed to be Markov correlated quantum noise. Mathematically, memory of
channels is incorporated using the technique of Macchiavello and Palma \cite%
{Chiara-1}. The results obtained show that memory of the channel increases
the classical capacity of the channels by considerable amount. The
comparison of classical capacity and entanglement-assisted classical
capacity of depolarizing channel, given in Fig. 1, shows that prior
entanglement is advantageous for the transmission of classical information
over quantum channels as compared to coding with entangled states.

\end{subequations}

\end{document}